\documentclass[aps]{revtex4}

\usepackage{graphicx}
\usepackage{dcolumn}
\usepackage{bm}
  
\begin{document}
\title{Weighted Density Approximation Description of Insulating YH$_3$
and LaH$_3$}
\author{Zhigang Wu}
\affiliation{Geophysical Laboratory, Carnegie Institution of Washington,
5251 Broad Branch Road, NW, Washington, DC 20015}
\author{R.E. Cohen}
\affiliation{Geophysical Laboratory, Carnegie Institution of Washington,
5251 Broad Branch Road, NW, Washington, DC 20015}
\author{D.J. Singh}
\affiliation{Center for Computational Materials Science,
Naval Research Laboratory, Washington, DC 20375}
\author{R. Gupta}
\affiliation{Commissariat a l'Energie Atomique, Centre d'Etudes de Saclay, \\
Service de Recherches de M\'{e}tallurgie Physique Commissariat \`{a} l'Energie Atomique,\\
91191 Gif Sur Yvette Cedex, France}
\author{M. Gupta}
\affiliation{Institut des Sciences des Materiaux, Batiment 415, \\
Universite Paris-Sud, 91405 Orsay, France}
\date{\today}

\begin{abstract}

Density functional calculations within the weighted density approximation (WDA)
are presented for YH$_3$ and LaH$_3$. We investigate some commonly used  
pair-distribution functions G. These calculations show that within a 
consistent density functional framework a substantial insulating gap can be 
obtained while at the same time retaining structural properties in accord with
experimental data. Our WDA band structures agree with those of $GW$
approximation very well, but the calculated band gaps are still 1.0-2.0 eV
smaller than experimental findings.

\end{abstract}
\maketitle


The rare earth trihydrides, $M$H$_3$, where $M$ = Y, La -- Lu,
are the subject of considerable practical and theoretical interest. The rare 
earths absorb hydrogen readily and in the trihydride form have among the largest 
hydrogen to metal ratios of the elemental hydrides \cite{libowitz}.
The fully hydrogenated materials are insulating, while those with less 
hydrogen are metallic, and in thin film form they can be 
used as switchable mirrors \cite{huiberts}. It is important to demonstrate 
a correct description of these trihydrides in first-principles approaches in 
order to validate calculations for more complex metal hydrogen systems. 
However the standard first-principles calculations based on density functional 
theory (DFT) within the local density (LDA) and generalized gradient (GGA) 
approximations incorrectly predict metallic behavior for the rare earth 
trihydrides. Here we explore application of the weighted 
density approximation (WDA). We find a WDA that works for these materials 
(and others), indicating that the gap problem here is not due to unusual 
correlations or quasiparticle corrections, but is simply a 
problem with the LDA and GGA exchange correlation (xc) functionals. 

The rare earth dihydrides, $M$H$_2$, are metallic. The trihydrides on the 
other hand are insulating with optical gaps of the order of 2 eV, with a 
metal insulator transition at a hydrogen content correlated with rare earth ionic
radius \cite{huiberts}. This insulating gap of the trihydrides
may be understood in terms of simple picture, in which H forms a negative ion, 
H$^-$, fully ionizing the metal atoms, thus forming a gap between H bands 
and higher lying metal $sd$ conduction bands. This simple picture is supported 
by infrared and photoemission measurements \cite{rode,osterwalder}.
It is consistent with the fact that all the rare earth trihydrides are
insulators despite differences in the details of their crystal structures.
These range from the $fcc$ derived structure of LaH$_3$ to the not fully 
solved hexagonal structure of YH$_3$ \cite{miron,pebler,udovic,zogal}.

Computations within the LDA predict a metallic band structure with
substantial overlap between H and metal derived bands \cite{zogal,swit,dekker,wang}. 
This severe gap problem could be due to several possible causes: (1) use 
of an incorrect crystal structure; (2) an extreme case of LDA error in predicting gaps; 
(3) unusual correlation effects; or (4) self-interaction errors in LDA.   
Early attempts at resolving this discrepancy focused on crystal structure. Indeed 
it was shown that with optimized crystal structures, DFT
calculations could yield insulating band structures for YH$_3$
\cite{kelly,ahuja}. Unfortunately, these relaxed structures are not 
consistent with diffraction measurements, and such an explanation would
not explain the similar insulating properties of the other
$M$H$_3$ compounds. 

DFT is a theory of ground state properties, and the eigenvalues of the 
Kohn-Sham equations cannot be equated to quasiparticle excitations \cite{onida}. 
Nevertheless, any non-pathological accurate approximation to the
exact DFT must predict an insulating Kohn-Sham band structure, because 
static responses and other ground state properties depend on the presence
or absence of a gap. For example, LO-TO splitting would be present in 
the phonon spectrum of ionic insulator like YH$_3$, but not in a metal.  

Recent studies on rare earth trihydrides have considered electron correlation effects
\cite{ng,eder,vg,vg2,miyake}, particularly electron correlations on the H sites. 
However, since there are no partially occupied $d$ or $f$
orbitals in LaH$_3$ and YH$_3$, nor is there evidence that these materials
are near a quantum critical point, these hydrides would represent
a rather novel class of strong correlated materials.
Calculations done within the $GW$ approximation show that band gaps that are 
reconcilable with experiment are obtained, both with the cubic BiF$_3$
structure (LaH$_3$) and the hexagonal HoD$_3$ structure (a simplified YH$_3$ structure) 
\cite{vg,vg2,miyake}. The $GW$ method approaches the problem as being most
simply viewed as due to the treatment of excited states, but if the problem is 
due to self-interaction, it is fundamentally a ground state problem. The LDA makes 
large errors in description of the one-electron hydrogen atom because of incomplete 
cancellation of the Hartree and xc self-interactions. 
     One way to remove the spurious self interactions in LDA and GGA
     calculations is the so-called self interaction correction \cite{zunger}. In this
     approach an orbital dependent self interaction is calculated at
     each step and used to create an orbital dependent single particle
     Hamilitonian. This method suffers from basis set dependence, e.g.
     the computed correction is zero for a Bloch state representation.
We explore whether one can remain within conventional ground state DFT and 
obtain the proper electronic structure of rare earth trihydrides. 

The weighted density approximation (WDA) \cite{gunna76,alonso78} has 
the exact non-local form of DFT expression of xc energy, 
\begin{equation}
E_{\rm xc}[n] = \int\int \frac{n({\bf r})n({\bf r}^{\prime})}{|{\bf r-r}^{\prime}|}
[g_{xc}({\bf r},{\bf r}^{\prime})-1] d{\bf r} d{\bf r}^{\prime}.
\label{eq1}
\end{equation}
Since the exact pair-distribution function $g_{xc}({\bf r},{\bf r}^{\prime})$ is
elusive, a model WDA function $G$ is made,
\begin{equation}
g_{xc}({\bf r},{\bf r}^{\prime})-1=G[|{\bf r-r}^{\prime}|, \bar{n}({\bf r})],
\label{eq0}
\end{equation}
where the weighted density $\bar{n}$ can be determined from the xc hole condition:
\begin{equation}
\int n({\bf r}^{\prime}) G[|{\bf r-r}^{\prime}|, \bar{n}({\bf r})]
d{\bf r}^{\prime} = -1.
\label{eq2}
\end{equation}
This assures no self-interaction in a one-electron system. The WDA also gives the 
correct result for the uniform electron gas because it reduces to the LDA in this case.
A seemingly natural choice of $G$ would be of a uniform system, but Mazin and 
Singh \cite{mazin98} emphasized, based on the xc kernel, that $G$ of the uniform
electron gas is not optimal. People advanced several model forms of $G$,  
such as the Gunnarsson-Jones (G-J) \cite{gunna80} ansatz, 
$G^{\rm G-J}(r,n) = c\{1-{\rm exp}(-[\frac{r}{\lambda}]^{-k})\}$, with $k=5$;
and the Gritsenko {\it et al. } (GRBA) \cite{grit93} ansatz,
$G^{\rm GRBA}(r,n) = c \cdot {\rm exp}(-[\frac{r}{\lambda}]^{k})$, with $k=1.5$.
In this letter, we also used a homogeneous type of $G$ \cite{ggp02}, and another G-J 
type for $k=4$ (G-J4), which was tested by Rushton {\it et al. } for silicon 
\cite{rushton}. To avoid the unphysical WDA inter-shell interaction, we used the 
shell-partition method \cite{gunna76,singh93,wu2003}.

As mentioned before, although the exact structure of hexagonal YH$_3$ is still under
investigation, its insulating property is believed to be an 
electronic property, i.e., not sensitive to the crystal structure \cite{gogh,molen}. 
So we studied a simplified hexagonal LaF$_3$ structure with a 
Y$_2$H$_6$ unit cell \cite{wang}, in which Y atoms form a hexagonal close-packed
(hcp) structure. The measured lattice constants are $a=3.672$
\AA \ and $c/a=1.81$. We optimized the H positions using LDA and WDA (with four
types of $G$), and all the results are consistent with the previous LDA
result \cite{wang}. We also optimized the equilibrium volume with the calculated 
atomic positions and fixed $c/a$ ratio. Table \ref{tab1} shows that the WDA  
volumes are only about 2-3\% deviated from experimental data.

\begin{figure}
\includegraphics[width=0.8\textwidth]{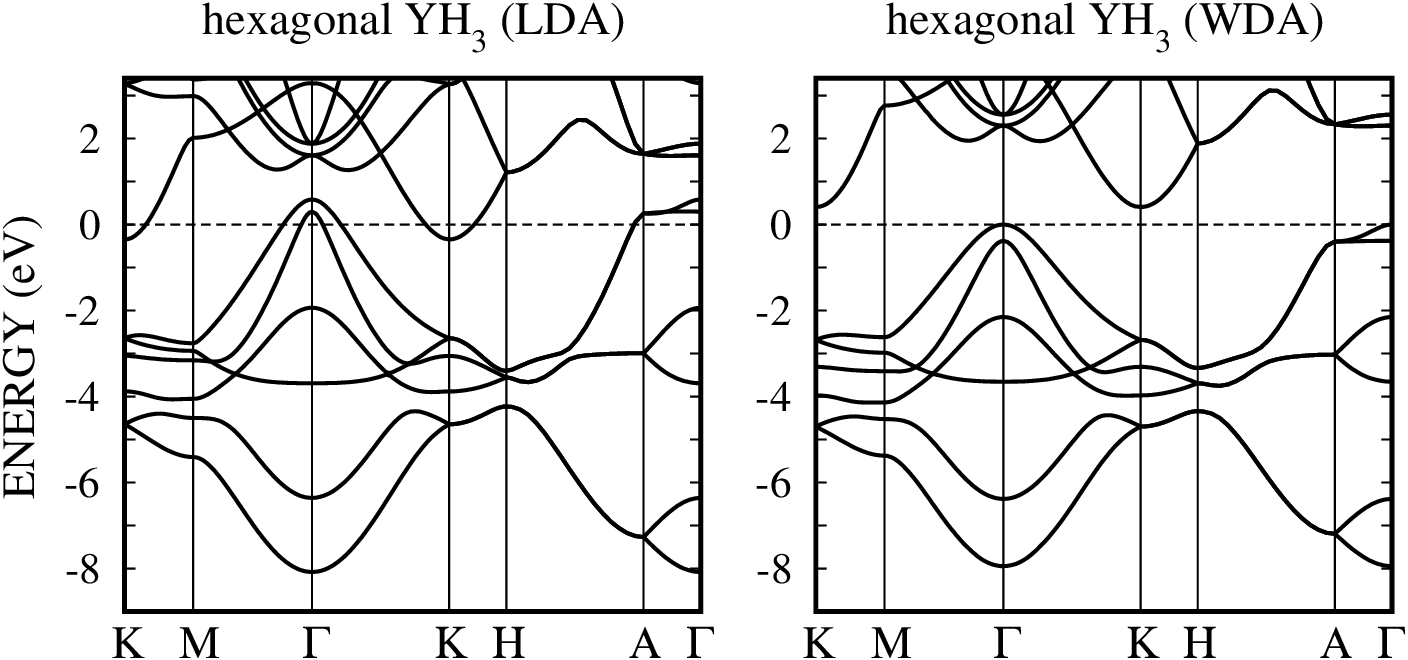}
\caption{\label{fig1} LDA and WDA (with the G-J4 type of $G$) 
electron bands of the hexagonal YH$_3$.}
\end{figure}

To determine the band structure of the hexagonal YH$_3$, We used the 
experimental volume and a $12 \times 12 \times 8$ special ${\bf k}$-point 
mesh. The calculated bands of the LDA and WDA (with the G-J4 type of $G$) are 
shown in Fig. \ref{fig1}. The LDA predicts a direct band gap of 1.0 eV at $\Gamma$ 
and an overlap of 0.93 eV between $\Gamma$ and K, which leads to a semi-metal. Our 
present LDA results are a little different from the previous (0.6 eV for the
$\Gamma$ gap and 1.3 eV for the overlap \cite{wang}) because we used the 
Hedin-Lundqvist xc function, whereas they used the Wigner ansatz. On the other 
hand, the WDA predicts a direct $\Gamma$ gap of 2.2 eV, and opens up a fundamental 
gap of 0.41 eV between $\Gamma$ and K. Thus the WDA predicts a substantial 
insulating gap. Our WDA band gaps are pretty close to the $GW$ calculations 
\cite{vg} of 0.6 eV for the fundamental gap and 2.9 eV for the direct $\Gamma$ gap. 
In the Brillouin zone of the more complicated tripled HoD$_3$ structure of 
YH$_3$, this indirect $\Gamma$-K gap will be folded to form a direct gap at 
$\Gamma$ point, and they are roughly of the same order of magnitude \cite{wang,vg}.

\begin{figure}
\includegraphics[width=0.8\textwidth]{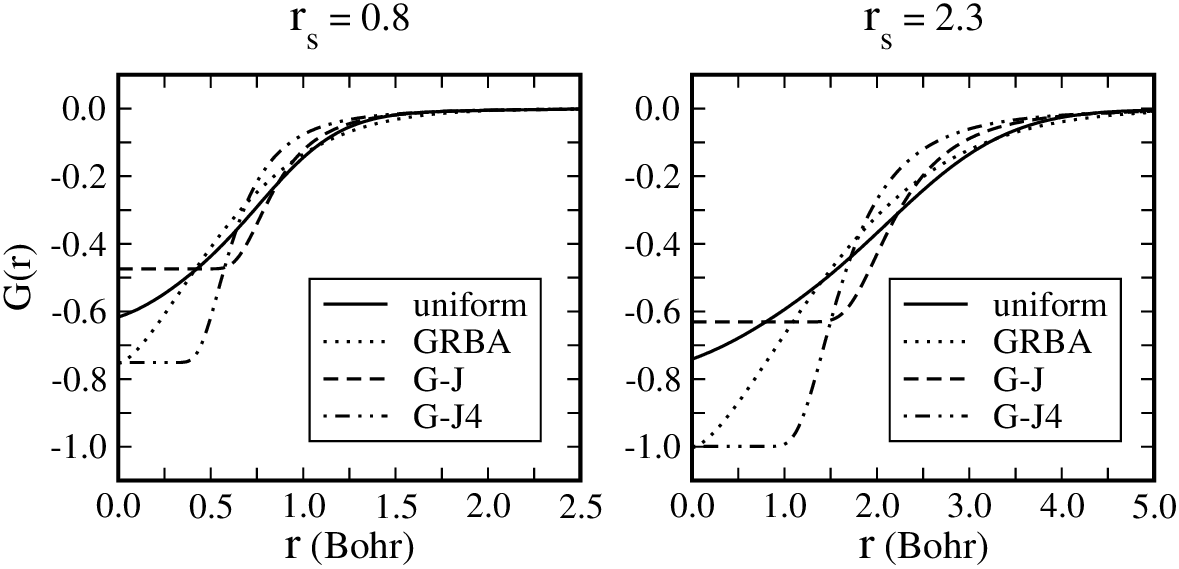}
\caption{\label{fig2} Four types of the pair-distribution function $G(r,n)$ 
with $r_s=0.8$ and $r_s=2.3$.}
\end{figure}

We also calculated the WDA bands with the other three types of $G$, and
the results are shown in Table \ref{tab2}. We found the WDA with all these $G$ 
functions gives better band gaps than the LDA. 
Except for the G-J4 form, the GRBA \cite{grit93} form is superior to the 
other two, because it predicts a smaller band overlap between 
$\Gamma$ and K. These results illustrate that the WDA band 
structure is sensitive to the choice of $G$. We compare these 
pair-distribution functions by drawing them together in Fig. \ref{fig2}, where
$n=3/(4\pi r_s^3)$, and in the WDA calculations of YH$_3$, $0.8 < r_s < 2.3$. 
We found the G-J4 form has the shortest range, and the range of GRBA ansatz is 
shorter than the other two. Our calculations indicate that the WDA with 
shorter range $G$ gives larger band gaps. Interestingly, such
functions, particularly GRBA, yield static responses closer to
Monte Carlo results for the uniform electron gas \cite{mazin98}. 

We performed LDA and WDA (with G-J4 only) calculations on cubic BiF$_3$ structure 
of YH$_3$ and LaH$_3$, in which the Y or La atoms constitute a face-centered cubic 
(fcc) lattice and H atoms are located at the tetrahedral and octahedral sites. 
Recently the insulating fcc YH$_3$ was stabilized by MgH$_2$ \cite{molen}. There is a large
discrepancy of the measured volumes of fcc YH$_3$. Ahuja {\it et al. } \cite{ahuja}
expected 32.2 \AA$^3$, while van der Molen {\it et al. } \cite{molen} extrapolated 
36.9 \AA$^3$ and Gogh {\it et al. } \cite{gogh} obtained 38.0 \AA$^3$. Our LDA and 
WDA equilibrium volumes are 32.4 \AA$^3$ and 35.2 \AA$^3$ respectively.
We used the experimental volume of 36.9 \AA$^3$. For LaH$_3$ the measured volume 
44.2 \AA$^3$ is close to LDA and WDA equilibrium volumes. We used a 
$12 \times 12 \times 12$ special ${\bf k}$-point mesh.

\begin{figure}
\includegraphics[width=0.8\textwidth]{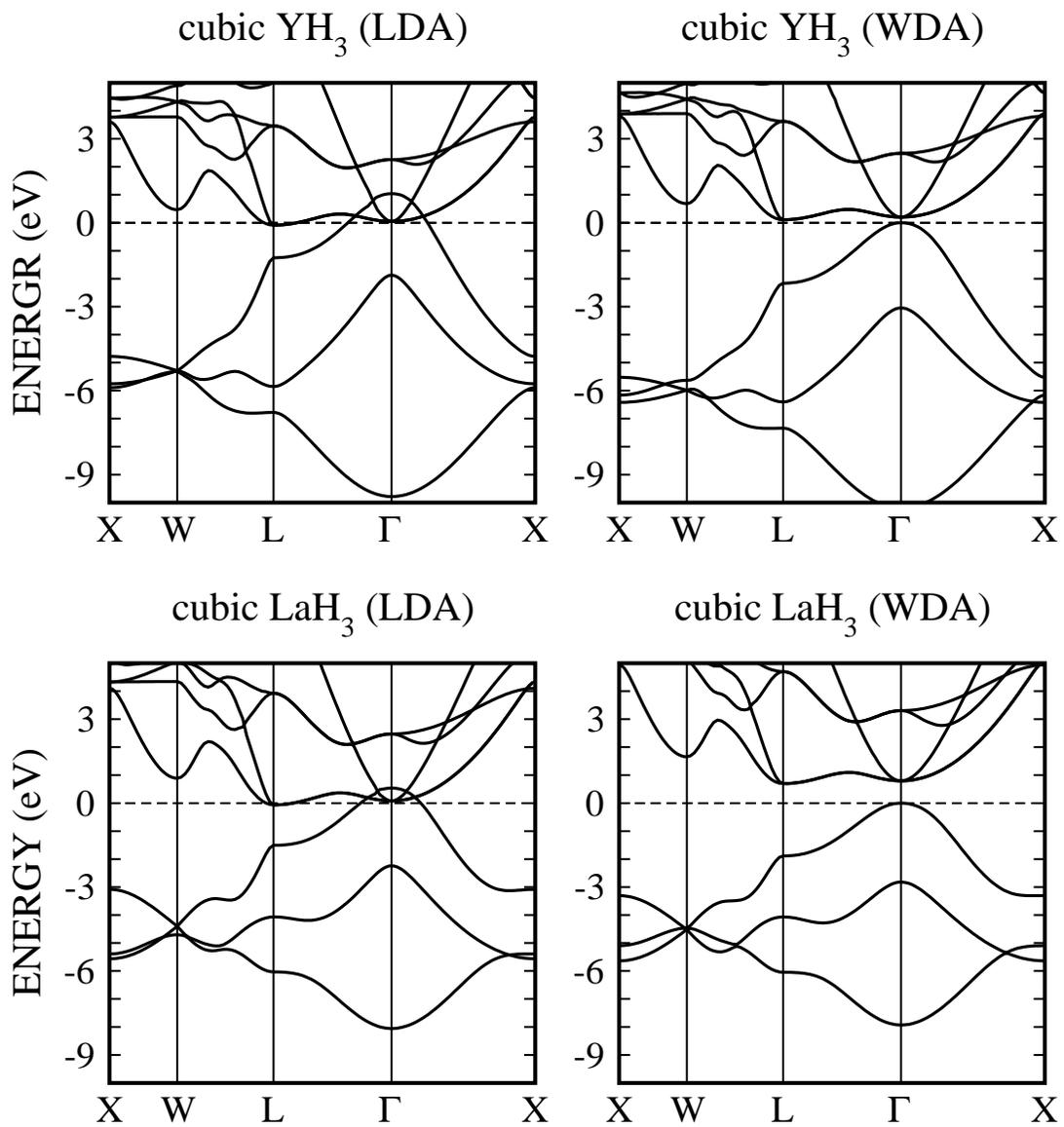}
\caption{\label{fig3} LDA and WDA (with the G-J4 type of $G$) 
electron bands of the cubic YH$_3$ and LaH$_3$.}
\end{figure}

In Fig. \ref{fig3} are the calculated band structures of cubic YH$_3$ and 
LaH$_3$. The LDA predicts that both of them semi-metals 
with direct overlaps of 1.0 eV for YH$_3$ and 0.46 eV for LaH$_3$
at $\Gamma$ point, which agree with previous calculations \cite{miyake,alford}. 
Contrary to the LDA, the WDA predicts that both trihydrides are semiconductors with 
direct $\Gamma$ gaps of 0.19 eV and 0.80 eV for YH$_3$ and LaH$_3$ respectively.
However the fundamental gaps are about 0.1 eV smaller, because the lowest 
conduction bands at L is slightly lower than that at $\Gamma$. Our WDA band gaps 
are in good agreement with previous $GW$ calculations. Alford {\it et al. } \cite{alford}
predicted a $\Gamma$ gap of 0.2-0.3 eV for YH$_3$ and 0.8-0.9 eV for 
LaH$_3$, and Chang {\it et al.} \cite{cbl} obtained 0.8 eV for LaH$_3$. We found that 
the hexagonal and cubic YH$_3$ have similar small gaps, suggesting that the insulating 
properties are of electronic origin rather than structural.

Another interesting feature of the WDA band structures is that compared
with the LDA, the overall valence bandwidth hardly changes. 
This is also consistent with previous $GW$ calculations 
\cite{vg,miyake,alford}, but in contrast to the many-body model Hamiltonian 
calculations which show a large valence bandwidth decrease \cite{ng}. It confirms 
the previous conclusion \cite{vg,alford} that rear-earth trihydrides are just 
band insulators, and that strong correlations are not required.

Although our DFT-WDA band gaps of YH$_3$ and LaH$_3$ agree with previous $GW$ 
quasiparticle calculations very well, the measured optical gap is 
$\sim$2.8 eV for YH$_3$ and $\sim$1.8 eV for LaH$_3$, which are still 
1-2 eV larger. Gelderen {\it et al. } \cite{vg} argued that the fundamental 
band gap of hexagonal YH$_3$ is not the measured optical gap due to the forbidden 
transition between the highest valence and the lowest conduction bands, and they 
concluded that the measured value corresponds to the direct gap at $\Gamma$ 
of the hexagonal LaF$_3$ structure. However the real hexagonal YH$_3$
is more complex and it may have low symmetry so that the dipole matrix
element is not zero. Our DFT-WDA method may underestimate band gaps 
due to the DFT gap discontinuity at the fermi level \cite{godby86}.

In conclusion, we have studied the band structures of the hexagonal YH$_3$ 
and cubic YH$_3$ and LaH$_3$ using the first-principles WDA method. The calculations 
showed that YH$_3$ and LaH$_3$ are both semiconductors, and this insulating property 
does not rely on the detailed crystal structure. The calculated band gaps agree with 
previous $GW$ calculations very well, and they are also consistent with experiment. 
Our WDA calculations also predicted good ground-state structures. 
The success of WDA may be attributed to the absence of self-interaction in H $1s$ sates.

We are grateful for helpful discussions with I. Geibels, P. Vajda and K. Yvon.
This work was supported by the Office of Naval Research under ONR Grants 
N000149710052, N00014-02-1-0506 and N0001403WX20028. Calculations were done on the 
Center for Piezoelectrics by Design (CPD) computer facilities and on the Cray 
SV1 at the Geophysical Laboratory supported by NSF EAR-9975753, and the 
W.\ M.\ Keck Foundation. We also thank the Institut du Developpement et des Ressources en 
Informatique Scientifique (IDRIS) for a grant of computer time. DJS is grateful 
for the hospitality of the University of Paris Orsay during part of this work.

\pagebreak

\begin{table}
\caption{With the calculated atomic positions and fixed $c/a$ ratio, the optimized 
equilibrium volumes (in \AA$^3$) of hexagonal YH$_3$ (Y$_2$H$_6$). 
Here a, b, c and d denote uniform, GRBA, G-J and G-J4 forms of $G$ respectively.
Numbers in parentheses are the percentage deviations from experiment. 
\label{tab1}}
\begin{ruledtabular}
\begin{tabular}{ccccc|c}
       LDA       & WDA (a) & WDA (b) & WDA (c)  & WDA (d) & Expt. \\  
\hline			
     72.8   & 75.0   & 77.4   & 75.6   & 79.4   & 77.6 \\
     (-6.2) & (-3.4) & (-0.3) & (-2.6) & (+2.3) &      \\
\end{tabular}
\end{ruledtabular}
\end{table}

\begin{table}
\caption{
Calculated LDA and WDA (a, b, c and d as seen in Table \ref{tab1}) fundamental and 
$\Gamma$ gaps (in eV), compared with $GW$ results \cite{vg}. 
A negative band gap means band overlap.
\label{tab2}}
\begin{ruledtabular}
\begin{tabular}{l|ccccc|c}
     &  LDA & WDA (a) & WDA (b) & WDA (c) & WDA (d) & $GW$   \\  
\hline			
     fd. gap      & -0.93  & -0.74  & -0.35  & -0.72  & 0.41  & 0.6 \\
     $\Gamma$ gap &  1.0   &  1.2   &  1.5   &  1.2   & 2.2   & 2.9 \\
\end{tabular}
\end{ruledtabular}
\end{table}

\end{document}